\documentclass[12pt]{article}
\usepackage{epsfig}

\def \be  {\begin{equation}}
\def \ee  {\end{equation}}
\def \ba  {\begin{eqnarray}}
\def \ea  {\end{eqnarray}}
\def \baa {\begin{eqnarray*}}
\def \eaa {\end{eqnarray*}}
\def \bb  {\begin{thebibliography}}  
\def \eb  {\end{thebibliography}}

\def \lab #1 {\label{#1}}

\def \fracs #1#2 {\mbox{\small $\frac{#1}{#2}$}}

\def \bin #1#2 {{\left({#1}\atop{#2}\right)}}
\def \as {\relax\ifmmode\alpha_s\else{$\alpha_s${ }}\fi}

\def \al #1 {\frac {\as({#1})}{\pi} }
\def \ds #1 {\ooalign{$\hfil/\hfil$\crcr$#1$}}

\newcommand \bea{\begin{eqnarray}}
\newcommand \eea{\end{eqnarray}}

\def\ra {\rightarrow}
\def\hepph  #1 {{\tt hep-ph/#1}}

\begin{document}

\begin{titlepage}
\renewcommand{\thefootnote}{\fnsymbol{footnote}}
\begin{flushright}
YITP-SB-08-23
     \end{flushright}
\par \vspace{10mm}
\begin{center}
{\large \bf
Some Basic Concepts of Perturbative QCD
\footnote{Based in large part on lectures presented at
the {\it School on QCD, Low-$x$, Saturation, and Diffraction};
Copanello (Calabria, Italy), July 1-14, 2007.  The author expresses thanks to
the organizers for the invitation and support.
This work was supported by the National Science Foundation, 
awards PHY-0354776, PHY-0354822 and PHY-0653342}
}

\end{center}
\par \vspace{2mm}
\begin{center}
{\bf George Sterman}\\
\vspace{5mm}
C.N.\ Yang Institute for Theoretical Physics,
Stony Brook University \\
Stony Brook, New York 11794 -- 3840, U.S.A.\\
\end{center}

\begin{abstract}
This is a brief review of some of the basic concepts of perturbative QCD, 
including infrared safety and factorization, 
relating them to more familiar ideas from quantum mechanics and relativity.
It is intended to offer perspective on methods and terms
whose use is commonplace, but whose physical origins are sometimes obscure.
\end{abstract}
  
  \end{titlepage}
  
\section{Asymptotic freedom in QCD}

We begin with a short portrait of
 quantum chromodynamics,   the unbroken, nonabelian gauge theory SU(3).
 The classical Lagrange density of QCD can be represented schematically by \cite{FGL}
\bea
{\cal L}_{QCD} = \sum_{q} \bar q\, \left( i\rlap{/}\partial-g\rlap{/}A +m_q\, \right)q - 
{1\over 4}\,  G_{\mu\nu,a}[A]\, G^{\mu\nu}_a[A]\, ,
\eea
where $q$ labels quark fields,
of mass $m_q$, and where
$G_{\mu\nu,a}[A]$ is the nonabelian field strength, including 
the self-couplings of the gluon field.
We can think of this expression as an analogy to
electrodynamics, the sum of kinetic terms for
the quarks and gluons, supplemented by various local interactions.
QCD is the Yang-Mills gauge theory \cite{YM} of quarks and gluons, in which
gluons are like photons with charge, so that the gluon field is a source for itself.
This nonlinearity, of course, is part of what makes QCD, and
the strong interactions it describes, the source of such varied phenomena.
It was realized early on that the quarks of QCD provide
just the right currents to couple to electromagnetic and
weak interactions, so that previous results based on the
analysis of those currents (``current algebra") could be taken over essentially
unchanged \cite{W73}.   In addition, this theory has just the right kind of forces: 
the QCD charge is ``antishielded", growing larger with increasing distances over which 
it is measured.  This is its famous property of asymptotic freedom \cite{AF}.

\begin{figure}[h]
\centerline{
{{\epsfxsize=9.5cm  \epsffile{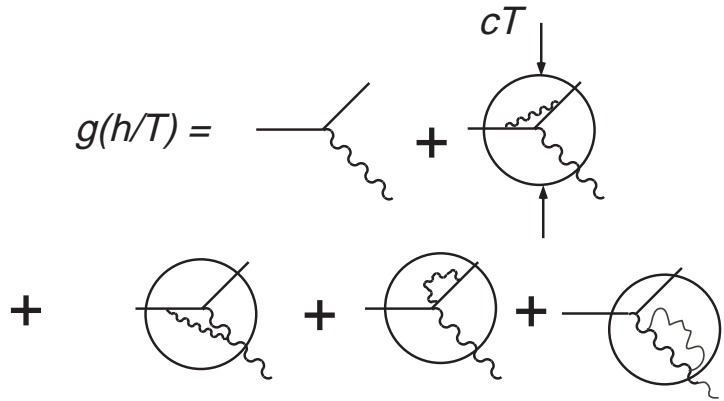}}}}
\caption{The running coupling defined by a sphere of radius $cT$.
\label{RngCpg}}
\end{figure}

Let us sketch how the asymptotic freedom of QCD
is established.   Working conceptually, imagine that we
{\it define} the strong coupling, $g(\hbar/T)$, as just  the amplitude 
for a quark to emit a gluon within a
sphere of radius $cT$, with $c$ the speed of light.   
So defined, $g(\hbar/T)$ is a {\em running coupling},
varying with the scale at which it is measured.
The concept is illustrated in Fig.\ \ref{RngCpg}.   In a sense, we
send a quark into the sphere, wait a time of order $T$,
and see if it comes out accompanied by precisely one gluon.
The amplitude for this to happen is given by an infinite set
of perturbative diagrams, each representing a particular set of quantum mechanical
histories.   We show some of the lowest order diagrams in Fig.\ \ref{RngCpg}.

Now the diagrams within the sphere are described by
integrals that do not converge, because 
there are simply too many states with one or more additional gluons
at very large energy.   Nevertheless, with a bit of work, we 
can compute the $T$-dependence of $g(\hbar/T)$, or
in more familiar notation, the $\mu\equiv \hbar/T$-dependence of $\alpha_s(\mu)\equiv
g^2(\mu)/4\pi$.  With
$n_f$ different (flavors of) quarks, we find
\bea
\as(\mu) \equiv {g^2(\mu)\over 4\pi} 
= 
 { \as(\mu_0) \over 1 + b_0\frac{\as(\mu_0)}{4\pi}\ln\left({ \mu^2\over\mu_0^2}\right)} \ 
\equiv
{ 4\pi \over b_0 \ln \left( \mu^2\, /\, \Lambda_{\rm QCD}^2\right)}\, ,
\label{alpharun}
\eea
where we adopt the common notation $b_0 = 11 -2n_f/3$,
and where $\alpha(\mu_0)$ and
the scale $\Lambda_{\rm QCD}$ can be thought of as integration
constants.   The value of $\Lambda_{\rm QCD}$ is set by any boundary
condition for $\alpha_s(\mu_0)$ at any scale $\mu_0$.   In other
words, it is set by nature, once we learn to measure
$\alpha_s$ at a given scale.
Eq.\ (\ref{alpharun}) expresses asymptotic freedom, according to which $\as(\mu\ra\infty) \to 0$. 
In QCD, the colors of virtual gluons  ``line up", like
neighboring magnets, a feature that depends on both the spin 
and the  self-interactions
of the gluons built into the Lagrangian.
The smaller the sphere, the fewer the lined-up magnets,
and the weaker the interaction.

An essential result of asymptotic freedom in 
QCD is that radiation becomes weaker as
momentum scales increase, or equivalently distances (like $cT$
above) decrease.   In effect, near a color source, the coupling constant is weak,
a feature that leads to the famous approximate ``scaling" observed
in deep-inelastic scattering, which we will describe below.
Correspondingly, far from a source, the coupling constant appears to grow,
a feature that is at least consistent with (although by no means 
ensuring) the observed confinement of colored quarks and gluons.

In the years leading up to the discovery of QCD, a template \cite{B69,BP,CHM}
had been developed to connect the behavior of the running
coupling in any field theory with what we now call parton distributions,
which we will denote as $f_{i/H}(\xi,\mu)$, 
for partons $i$ carrying momentum fraction $\xi$
of hadron $H$ \cite{PM}.   Here, $\mu$ is a renormalization
scale, very much analogous to $\hbar/T$ above, and can
be thought of as determining the scale at which we
probe hadron $H$ to count these partons $i$, leading
to probability density $f_{i/H}(\xi,\mu)$.
This probe-scale dependence is encoded in 
sets of  ``anomalous dimensions", which can be computed
as power series in the couplings,
\bea
\gamma_N(\alpha_s) = \frac{\as}{\pi}\, \gamma_N^{(1)}+ \dots\, ,
\eea
where for QCD, we know that 
$\as(\mu)$ vanishes as $\mu$ increases.
For moments of the parton distributions, 
$\bar f(N) = \int_0^1 d\xi \xi^{N-1}f(\xi)$,
 the general analysis gives \cite{B69,BP,CHM}
\begin{eqnarray}
{\bar f_{i/H}(N,\mu)} 
= {\bar f_{i/H}(N,\mu_0)}
\; \exp\left[\;  -{1\over 2}\; \int_{\mu_0^2}^{\mu^2}\;
{d\mu'{}^2\over \mu'{}^2}\ \gamma_N(\alpha_s(\mu'))\; \right]\, ,
\label{fNevol}
\eea
and with $\as(\mu) = 4\pi/b_0\ln(\mu^2/\Lambda_{\rm QCD}^2)$, we get:
\bea
{\bar f_{i/H}(N,Q)}\
= {\bar f_{i/H}(N,Q_0)}
\left({\ln(Q^2/\Lambda_{\rm QCD}^2)
\over 
\ln(Q_0^2/\Lambda_{\rm QCD}^2)} \right)^{-2\gamma_N^{(1)}/b_0 }\, .
\end{eqnarray}
Once the $\gamma_N$'s were computed
at one loop (and eventually all the way to three loops \cite{as3split}), it all worked.  

To get back to the parton distributions, $f_{i/H}(\xi,\mu)$, we can invert the moment transform.
We can then compute structure function $F_2(x,Q^2)$, which describes
deeply inelastic scattering in terms of the variable
$x=Q^2/2p\cdot q$ and the momentum
transfer $Q^2=-q^2>0$ (see Eq.\ (\ref{Tmunudef}) below).
The data of Fig.\ \ref{evolfig} shows exactly a pattern predicted by
the explicit forms of the $\gamma_N$'s:
approximate scaling ($Q$-independence) at moderate $x$
and pronounced evolution ($Q$-dependence) for small $x$.
Perfect scaling for the structure functions
would follow for vanishing coupling.   This corresponds to 
$\mu$-independent parton distributions, as in the {\it parton model},
which provided a successful description of the first moderate-$x$ data,
in which $Q$-dependence seemed weak if not absent altogether \cite{PM}. 
\begin{figure}[t]
\hbox{\hskip 0 in \epsfxsize= 5 cm \epsffile{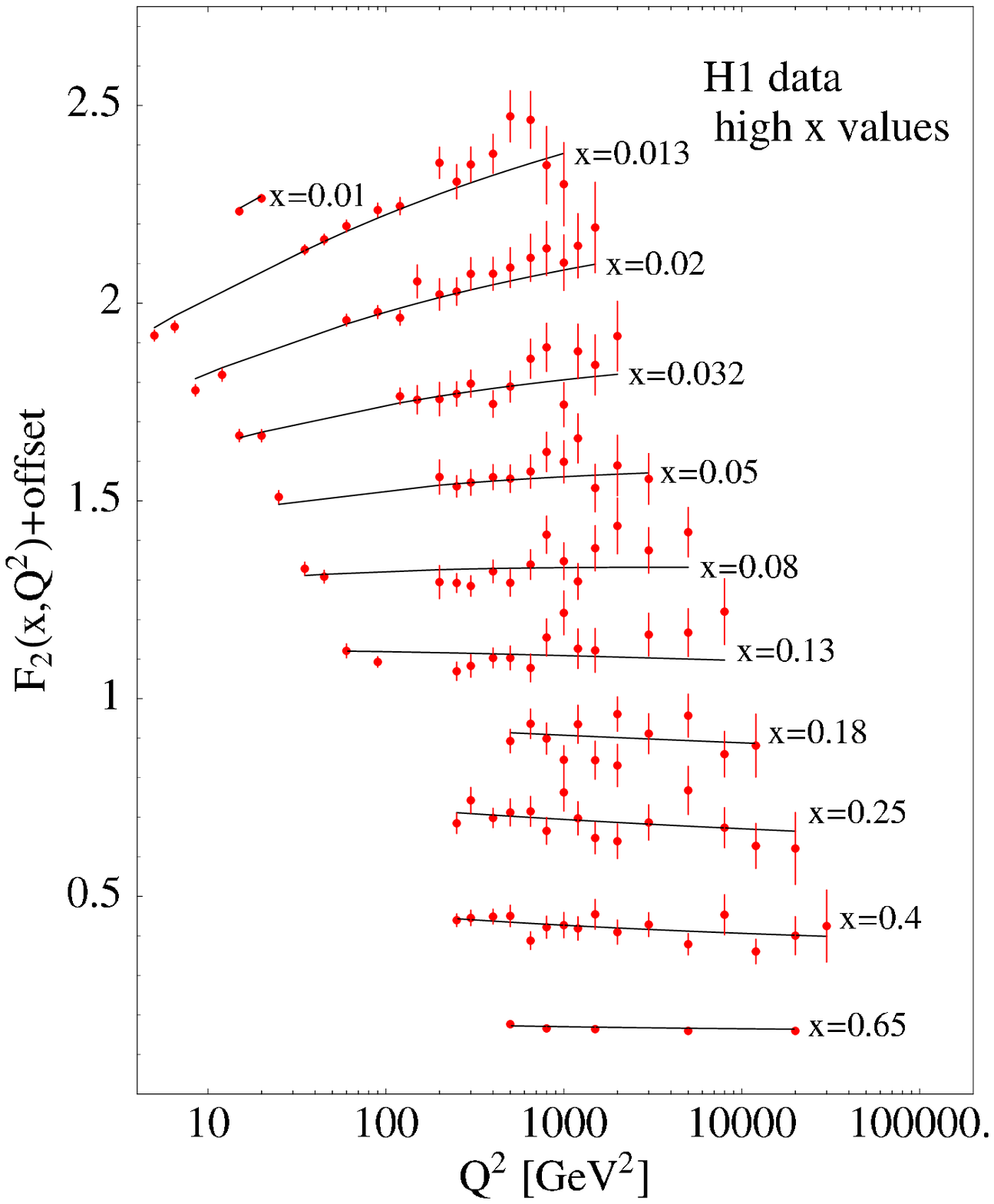}
\hskip  1 in \epsfxsize 5 cm \epsffile{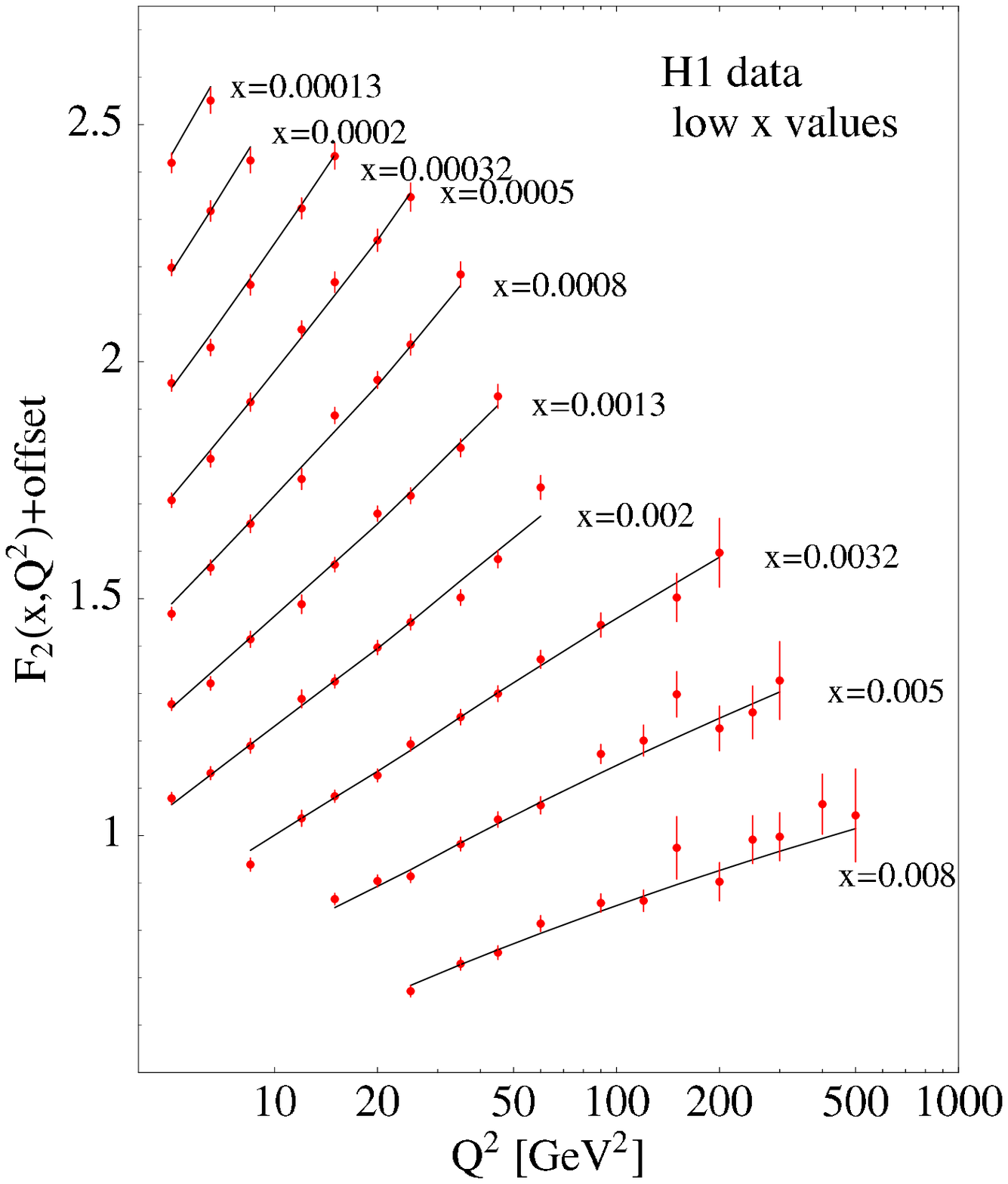}\\
}
\hbox{ \hskip 1 in (a) \hskip 3 in (b)}
\caption{(a) Approximate scaling at moderate $x$.
(b) Pronounced evolution for smaller $x$.  Data from
the H1 experiment at HERA.
\label{evolfig}}
\end{figure}

As has been widely recognized, the  asymptoic freedom 
of the QCD running coupling is a result of historic significance.
This is as much because it opens the door to new studies,
as because it explained previously mysterious features of
nature.   A tongue-in-cheek analogy that I like is
\bea
{\rm {Scaling \over QCD}\ =\ {Elliptical\ 
Orbits \over Newtonian\ Gravity}}\, .
\eea
In its explanation of approximate
scaling (and the violation, or ``breaking" of scaling), 
asymptotic freedom is a beginning, not an end.
For Newtonian gravity, the immediate challenge to 
the inverse-square law was the three-body problem
(moon-sun-earth, for example).   For QCD it is how to study
a theory in which the fundamental degrees of freedom
are masked by confinement.   
The ultimate goal might be expressed in a similar spirit as
\bea
\rm {Nuclear\ Physics \over QCD}\ =\ {Chemistry \over QED}\, .
\eea
A short summary of  questions we must ask in this context include: can we 

\begin{itemize}

\item{} Study the particles that give rise to electroweak currents (quarks)?

\item{} Study the particles that provide the forces (gluons)?

\item{} Expand in the number of gluons (i.e., use perturbation theory)?

\end{itemize}

In QCD the fundamental quanta are confined, 
and (at least in the absence of extreme temperature
and pressure) observed
hadrons are bound states.
The scattering of bound states confronts us
with the complex structure of these hadrons,
and the strong forces that hold them together
on length scales comparable to $1/\Lambda_{\rm QCD}$.
A question that was raised often in the early days
of QCD was quite simply, ``Does this make sense at all?"

\section{Learning to Calculate with the Theory}

Our first observation is that not all is hopeless.  Certain
quantities even in a confining theory 
are quite ``perturbation theory-friendly".  

\subsection{Correlations and the S-matrix}

The classic examples of quantities closely related to perturbation theory are
correlation functions between color-singlet currents at
short distances, schematically,
\bea
\langle N |\, J(z)\, J(0)\, | N\rangle &=& C_N\left(z\mu,\alpha_s(\mu)\right)
\nonumber\\
&=& C_N\left(1,\alpha_s(1/z)\right)\, ,
\label{JJexpect}
\eea
in some state $N$.   When $N$ is the vacuum,
the primary example is the total $\rm e^+e^-$ annihilation cross section,
for which $J$ is the electroweak current.  
In this case, the function $C_N$ can be expressed as a power
series in the coupling evaluated at the momentum scale of inverse distance $1/z$
(here treated as a simple scalar).
Any such quantity, which depends  only on one or more short distance scale,
is said to be {\it infrared safe}.
 When $|N\rangle$ is a nucleon state, such matrix elements are related to deep-inelastic
scattering.   In this case, the function $C_N$ is somewhat
more complex, and the matrix element is not itself infrared safe,
but its dependence on the short length scale
is still computable, using the factorization formalism that we review below.

Calculating an $S$-matrix element in perturbative QCD, however, is
pretty hopeless,
\bea
\langle B\ {\rm out}|A\ {\rm in}\rangle &=&
f\left(Q/\mu,m/\mu,\alpha_s(\mu)\right)
\nonumber\\
&=&
f\left(1,m/Q,\alpha_s(Q)\right)
\nonumber\\
&=&
f\left(Q/m,1,\alpha_s(m)\right)\, ,
\eea
where $A$ and $B$ are hadronic states, $Q$
is a hard scale, and $m$ denotes various 
soft scales in the theory, including the masses
of light quarks, the (perturbatively vanishing) mass\
of the gluon, and the strong-coupling scale $\Lambda_{\rm QCD}$
from Eq.\ (\ref{alpharun}).
No matter what choice we take for the scale in the running
coupling, we encounter large ratios of the energy to
fixed mass scales.
If S-matrix elements are not accessible, are we doomed to compute only correlations of currents?
The answer turns out to be ``not quite", and here we can turn to another strand
of the story.

\subsection{Structure of final states: Cosmic rays to quark pairs}

As it turns out, not being able to compute S-matrix elements is
not the same as begin {\it forbidden} to ``look inside the final state".
In fact, as we now know, it is possible to see in certain final states
a direct portrait of quarks and gluons in the form of ``jets" of nearly collinear
high-energy particles.  The story of the term ``jets" actually begins before QCD, in fact
even before the paper of Yang and Mills.   This
is the tale of particle jets in cosmic rays.

While tracing back some references, I was surprised to read
in a paper from 1957,  that
 ``The average transverse momentum
resulting from 
 our measurements is $p_T$=0.5 BeV/c
for pions [a table] gives a
 summary of
jet events observed to date \dots" 
\cite{Edwards}.  
Evidently, the jets associated with perturbative QCD did not by themselves
give rise to the term.   What was being reported
in this paper was a spray of particles of very high
energy  but limited transverse
momentum (a BeV is a GeV), observed in
cosmic ray events, as seen in emulsions.

Somewhat over ten years after Ref.\ \cite{Edwards}, accelerators had been
developed to study the multi-GeV annihilation of electrons and
positrons into virtual photons, which can then decay
into anything that carries charge.   In the meantime,
the quark model had been invented, and the quarks
carry charge.   So, what was going to happen?   If (as everyone
suspected) we wouldn't see the quarks because of
confinement, what would we see?   Jets?  In
{\it Physical Review}, Drell, Levy and Yan \cite{DLY} took
the step of extending the parton model from deep-inelastic
scattering to $\rm e^+e^-$ annihilation, and 
built into their model the same limited
transverse momentum that had been observed
a decade earlier in cosmic rays,
describing the limitation as a cutoff:  ``Because of our cutoff $k_{\rm max} \ll |q|$ \dots
The distribution   of secondaries in the colliding ring frame
will look like two jets \dots "

\begin{figure}[h]
\centerline{ {\epsfxsize=8cm  \epsffile{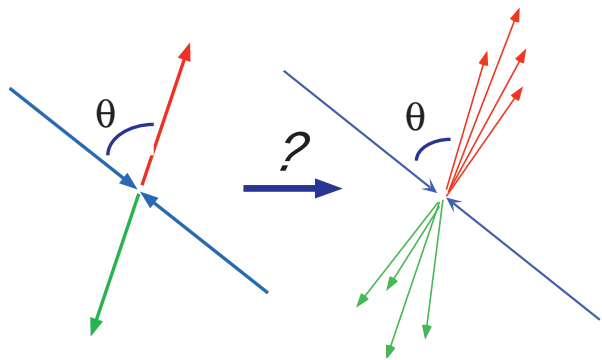}}}
\caption{Representation of the conjecture that there is a relation between
quark pairs and jets for $\rm e^+e^-$ annihilation. \label{jetdraw}}
\end{figure}

Now this was a real {\it prediction} for the nature
of final states in $\rm e^+e^-\rightarrow$ hadrons,
and following the spirit of the parton model, Drell, Levy and Yan suggested that the angular
distribution of the jets would follow the same angular distribution as a quark pair,
or any other spin-$1/2$ pair, in Born approximation,
$1+\cos^2\theta$, with $\theta$ the angle to the beam axis (see Fig.\ \ref{jetdraw}).   
 In this picture, partons ``fragment" into hadrons.
Whether this would happen was a question to ask of both nature and of QCD.
Would the final states look like this?

\begin{figure}[h]
\centerline{{\epsfig{file=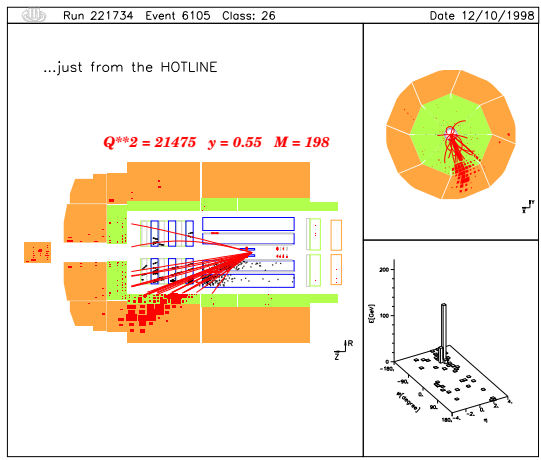,width=60mm,angle=90,clip=}}}
\caption{A jet in deep-inelastic scattering. Event recorded at
the H1 experiment at HERA.  \label{disjet}}
\end{figure}

\begin{figure}[h]
 \centerline{ \epsfxsize=7cm \epsffile{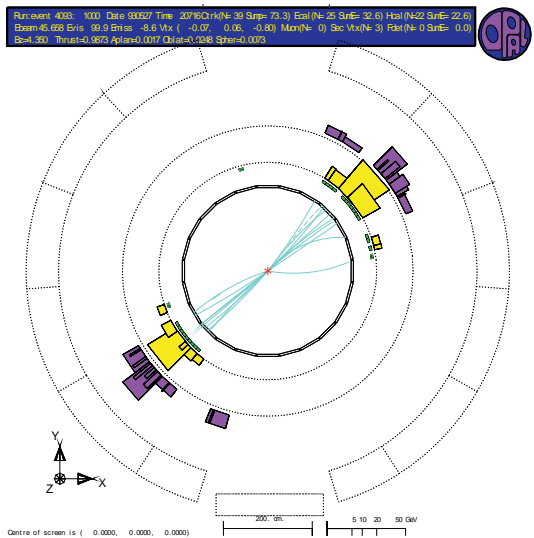}}
\caption{A jet pair in $\rm e^+e^-$ annihilation. 
Event recorded at the Opal experiment at LEP.  \label{epemjet}}
\end{figure}

\begin{figure}[h]
  \centerline{\epsfxsize=8cm \epsffile{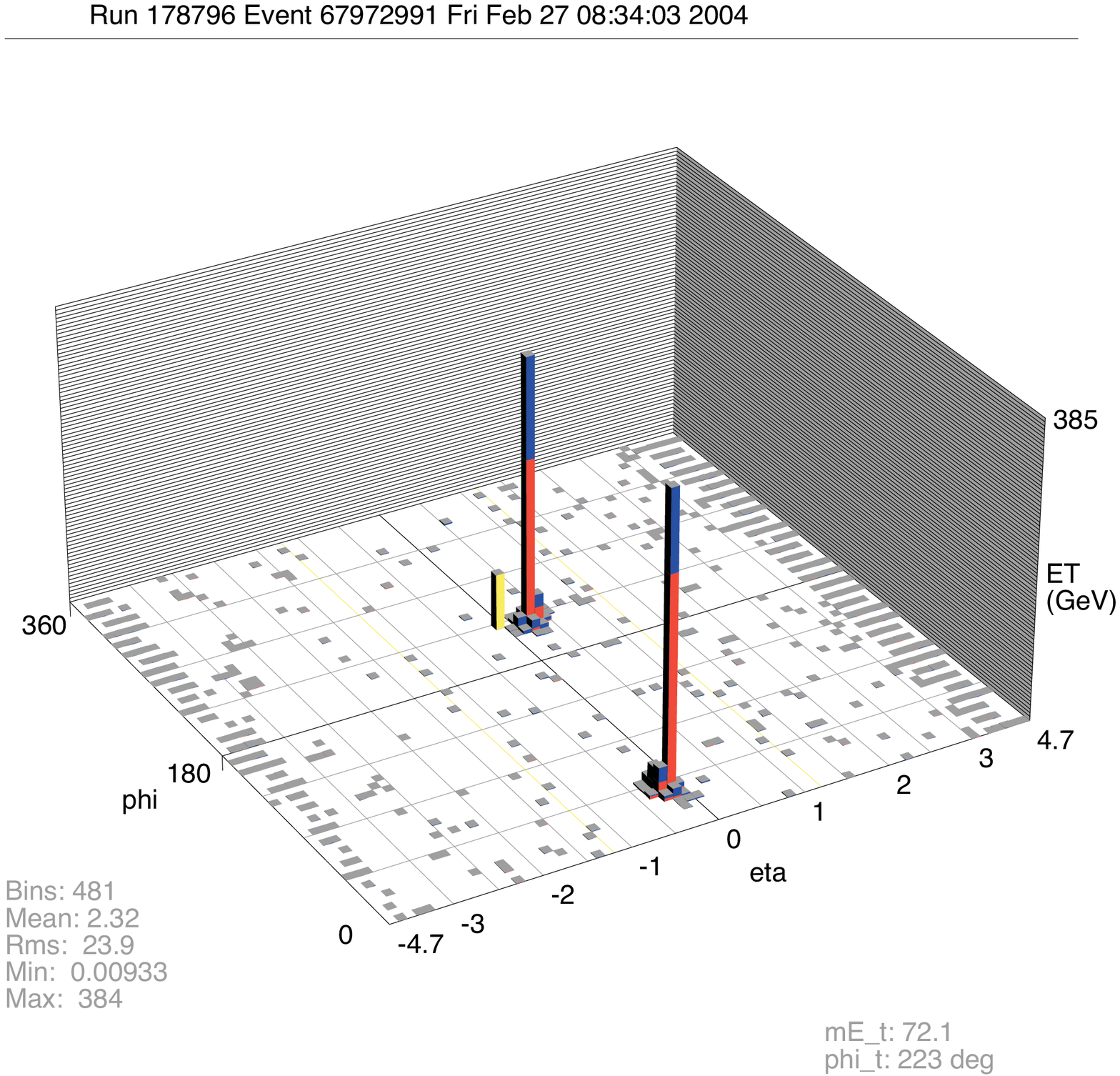}}
  \caption{Jet pair at a hadron collider. 
  Event recorded by the D0 experiment at Fermilab.  \label{hhjets}}
  \end{figure}

 In nature, they did, as shown by the analysis of
 Hanson et al.\ at SLAC in 1975 \cite{Hanson}.
   And, in the fullness of time, that's what happens in 
   deep-inelastic scattering, in $\rm e^+e^-$ annihilation
   and in hadron-hadron scattering.
Figures \ref{disjet}-\ref{hhjets} show nature's answer to the
question of whether jets exist.   
  
\subsection{How to calculate jet cross sections}

Clearly, 
we can observe the jets, but we still have to ask
whether we can calculate anything about them.
Here we can hope to benefit from the asymptotic freedom
of QCD, but as we've seen, we
have to be careful --  the S-matrix cannot be treated by
short-distance analysis alone.  If so, how can we hope
to compute cross sections?

We can get insight into the challenges involved, and
their possible solutions, by recalling the related ``infrared
problem" of QED, and its ``solution".  As is often the case, the
problem is related to asking the wrong question, and
the solution to identifying the right one.    

In QED, typical exclusive cross sections have infrared divergent corrections
in perturbation theory,
which show up as logarithmic dependence on the (vanishing) mass of the photon.
This happens 
as soon as we go to the order $\alpha_{\rm EM}\equiv e^2/4\pi$ correction of a Born cross section, say
in electron-electron scattering at momentum transfer $Q$:
\bea
\sigma^{(1)}_{{\rm ee\rightarrow ee}}\left(Q,m_e,\ {m_\gamma\ra 0},\alpha_{\rm EM}\right) 
\sim \alpha_{\rm EM}\; \beta_{{\rm ee\rightarrow ee}}(Q/m_e)\;  {\ln {Q\over m_\gamma}}\, ,
\eea
with $\beta_{{\rm ee\rightarrow ee}}(Q/m_e)$ a function that is finite for vanishing
photon mass, $m_\gamma$.
Following the famous Bloch-Nordsieck analysis \cite{BNYFS},  however,
we trace this divergence to asking an unphysical question,
the probability for one or more charged particles to scatter,
and in the process be accelerated, while emitting no radiation at all.
In effect, we are computing the probability of something that {\it never} happens.

The classical theory demands radiation, and classical radiation
requires an essentially unlimited number of very low-energy photons.
This is Bohr's correspondence principle between quantum and classical mechanics.
Rather than count the number of photons (zero, one \dots ), 
Bloch and Nordsieck \cite{BNYFS} counseled that we introduce an
{\it energy resolution},  $\epsilon Q$ with $\epsilon \ll 1$,
and then sum over final states with arbitrary photon emission, as long as
the total energy comes in below the energy
resolution.  Experimentally, this is not a choice, but a
necessity, because our apparatus will always
miss some photons if they are soft enough.   At first, however,
it sounds complicated.  How can we sum over {\it all}
soft photons?  But this will not be necessary.

Following the Bloch-Nordsieck procedure, suddenly
the full order $\alpha_{\rm EM}$ correction with an energy resolution becomes finite
by itself, the
log of $m_\gamma/Q$ being replaced by a log of $\epsilon$,
as the result of a cancellation between the final states with and without an extra photon.
As long as $\alpha_{\rm EM} \ln (1/\epsilon) \ll 1$ (which is almost inevitable),
the correction is small,
\bea
 \overline{\sigma}^{(1)}_{{\rm ee\rightarrow ee}+X(\epsilon)}\left(Q,m_e,{\epsilon Q},\alpha_{\rm 
EM}\right) 
\sim \alpha_{\rm EM}\; \beta_{{\rm ee\rightarrow ee}}(Q/m_e)\; {\ln {1\over \epsilon}}\, .
\eea
The magic (and beauty) of this is that we don't have to sum over an infinite number
of soft photons, even though this is the root cause of the problem!

Now let's think about QED in the very high-energy limit.
On a closer look, we find that the function $\beta_{{\rm ee\rightarrow ee}}(Q/m_e)$
itself has a log of $Q/m_e$.   Given that the dominant term
depends only on the ratio,  we can as well trade the high-energy
limit for the double, photon-and-electron, zero mass limit.    
But in this case, our energy resolution is not enough to
produce finite cross sections.
If, however, we can solve this problem in QED, we may be able to
solve it as well in high-energy QCD, where the high-energy limit
also involves logarithmic enhancements in ratios of momenta
to all particle masses.

We wish to look for quantities that are
capable of measurement, have a single hard momentum scale, $Q$,
and which nevertheless have {\it no}
powers of $\ln(Q/m)$, only at worst $(m/Q)\ln(Q/m)$.
Such quantities become functions
of only that single hard scale, and are calculable
as a power series in the
coupling $\alpha_s(\mu)$, with $\mu=Q$,
without introducing any large ratios when $Q\rightarrow \infty$
with fixed masses, or equivalently $m\rightarrow 0$
at fixed $Q$.   These are quantities for which
asymptotic freedom can be naturally applied,
and are, in the terminology mentioned above, infrared safe.

We've already seen that an energy resolution alone is not enough
for infrared safety.
Progress can be made, however, by an analogy to the argument
for an energy resolution based on the unobservability of 
arbitrarily soft photons.   We can just as well say that two
exactly collinear massless particles cannot be 
distinguished from a single massless particle of the same total momentum
(and total quantum numbers), whether
that momentum is soft or not.  That is, if $p^2=0$ and $p'{}^2=0$
and if $p^\mu$ and $p'{}^\mu$ are collinear, then $(p+p')^2=0$ as well.
So whether the combination is a single particle or two particles
is not easy to distinguish.   

This approach works, and enables us to take the
zero-mass limit for all particles in QED {\it and} QCD.   Roughly speaking,
any cross section with an energy and an {\it angular}
resolution is infrared safe in $\rm e^+e^-$ annihilation \cite{SW}.
If two particles are closer together in direction than some
angle, $\delta$, then we treat them the same way as we do a single particle.
We'll also see that hadron-hadron cross sections of this sort,
while not themselves infrared safe,
contain an infrared safe factor that 
we can isolate.   

The conditions for infrared safety may also be
rephrased in a more general form as follows.
Any cross section that sums over all states that
(1) differ by the emission or aborption of soft particles,
or (2) by the splitting or recombination of exactly
collinear particles, is infrared safe \cite{SW,PDEFG,S78,Sterman:1994ce,
Brock:1993sz,Dasgupta:2003iq} (or contains an
infrared safe factor.)   It is worth noting that to prove infrared safety for
jet cross sections requires an extension of the beautiful theorems that
apply to fully inclusive transition probabilities \cite{KLN}.  This involves
a careful reanalysis of perturbation theory, and is especially dependent on
how the gauge invariance of the theory manifests
itself \cite{S78}.

The most direct application of infrared safety is to jet cross
sections in $\rm e^+e^-$ annihilation, exactly of the type
illustrated schematically by Fig.\ \ref{jetdraw} and in
experiment by Fig.\ \ref{epemjet}.   So long as the resolutions are
large, we can represent the use of asymptotic freedom for
jet cross sections by an appropriate choice of 
renormalization scale, taken here as the total energy $Q$,
\bea
\sigma\left(Q/\mu,\epsilon,\delta,\alpha_s(\mu)\right)
=
\sigma\left(1,\epsilon,\delta,\alpha_s(Q)\right)\, ,
\eea
where $\epsilon$ and $\delta$ represent the
energy and angular resolutions mentioned above,
or more generally other parameters that define 
the cross section.
Computed in this fashion, there is no need for
a transverse momentum cutoff of the sort 
invoked in Ref.\ \cite{DLY}.   The infrared safety of the
observable ensures that high-$p_T$ radiation
is suppressed by factors of $\alpha_s(p_T)$.
Such radiation is present, of course, but it
influences the infrared safe quantity through  
calculable corrections, just as the effects of soft
gluons influence QED cross sections in a finite
way at higher orders.

\subsection{The field-theoretic content of infrared safety}

Summarizing, we recount the ``sorrows" of QCD perturbation theory,
and how they can be overcome, at least in part.
First, there is color confinement, which may be 
interpreted as the statement that
matrix elements like
\bea
\int d^4x\, {\rm e}^{-ip\cdot x}\langle 0|\, T[ q_a(x) \dots ]\, |0\rangle\, ,
\eea
in which we take the Fourier transform of a quark or
other field with a nontrivial color representation
 has no
$p^2=m^2$ pole in a Green function, with $T$ time-ordering.  (This is confinement.)
 Second, poles at physical particle masses, such as $p^2=m_\pi^2$ for pions,
\bea
\int d^4x\, {\rm e}^{-ip\cdot x}\langle 0|\, T[ \pi(x) \dots ]\, |0\rangle\, ,
\eea
are not accessible to perturbation theory.

Despite all this, we are able to use infrared safety and asymptotic freedom
for such quantities as the total cross section for $\rm e^+e^-$
annihilation into hadrons.
What are we really calculating?  Totally inclusive examples like these are 
related by the optical theorem to forward-scattering amplitudes of the
general form
 \bea
 \int d^4x\, {\rm e}^{-iq\cdot x}\langle 0|\, T[  J(x) J(0) ]\, |0\rangle\, ,
 \eea
 involving color
singlet currents, just as in Eq.\ (\ref{JJexpect}).   Deep-inelastic scattering involves 
hadronic matrix elements rather than the ground state \cite{Sterman:1994ce,Brock:1993sz},
\bea
W_{\mu\nu} 
&=& 2\ {\rm Im}\,  {i\over 8\pi} \int d^4x\, e^{-i q \cdot x}  < N (p) |\; T[\;  
J_\mu (x)
J_\nu (0)\; ]\; | N (p)>
\nonumber\\
&=& -\left( g_{\mu\nu} - \frac{q_\mu q_\nu}{q^2}\right) \, F_1(x,Q^2)
\nonumber\\
&\ & \hspace{10mm}
+ \left(p_\mu - q_\mu \frac{p\cdot q}{q^2}\right)\left(p_\nu - q_\nu \frac{p\cdot q}{q^2}\right)\,
\frac{1}{p\cdot q}\, F_2(x,Q^2)\, ,
\label{Tmunudef}
\eea
for electroweak currents, $J_\mu$, and nucleon states, $|N(p)\rangle$,
with $x\equiv p\cdot q/Q^2$ and $q^2 = -Q^2 <0$.   The $F_i(x,Q^2)$ are
the same structure functions as shown
in Fig.\ \ref{evolfig}.
For such matrix elements, we will apply factorization properties, 
which will enable us to isolate the infrared safe factors referred to above.

 Another class of color singlet matrix elements enables
 us to describe jet-related cross sections \cite{Eflow}.
 These look like
\bea
\lim_{R\to \infty}\, R^2\, \int dx_0  \int d\hat n\, f(\hat n)\, {\rm e}^{-iq\cdot y}
\langle 0|\, J(0)  T[ \hat n_i T_{0i}(x_0,R\hat n) J(y) ]\, |0\rangle\, ,
\eea
with $T_{0i}$ the energy momentum tensor, and $\hat{n}$ a
vector on the unit sphere.   Such a matrix element
represents the action of a calorimetric detector, which measures
energy flow, and matrix elements such as these are
what we really calculate when we compute jet cross sections.
For a general cross section, we introduce a ``weight", given
by function $f(\hat{n})$.   
As long as all the derivatives,
$d^r f(\hat{n})/d\hat n^r$, of the weight are bounded, 
individual final states may have infrared divergences, but they                                                                                                                                                    cancel
in the sum over collinear splitting/merging and
 soft parton emission, because these transitions respect energy flow \cite{S78}. 
 We regularize the divergences dimensionally (typically)
and calculate the long-distance enhancements
in amplitudes,
only to cancel them in infrared safe cross sections.
 It is this intermediate step that makes many calculations
tough, and is part (not all) of why higher-order calculations
are so difficult.  It may be worth noting that 
one of the goals of a collider experiment is remarkably similar -- to control
late stage interactions of particles once they enter the detectors.

\section{Extracting Infrared Safety: Factorization}

Any cross section
with one or two hadrons in the initial state has inescapable
long-distance behavior, because a semi-inclusive sum over initial
states is simply not a practical option.   In effect, we can
choose the energy of the nucleon(s) that initiate our
scattering process, and sometimes their spin, but little else.
By construction, then, cross sections at hadronic colliders
are not infrared safe.
The technqiue of factorization, however, enables us to
isolate and extract infrared safe dependence in a large set of otherwise
long-distance phenomena. 
Here we review the physical basis of factorization,
and show how the factorization of a process also leads
to useful information on its energy-dependence,
including the evolution of the moments of
parton distributions, as in Eq.\ (\ref{fNevol}) above.

\subsection{Factorization}

The general form of a factorized
cross section (here multiplied by $Q^2$ to
make it dimensionless), is \cite{FactPrf,CSS89}
\bea
Q^2\sigma_{\rm phys}(Q,m)
=
{\omega_{\rm SD}(Q/\mu,\as(\mu))}\, \otimes\, f_{\rm LD}(\mu,m) + 
{\cal O}\left({1/ Q^p}\right)\, ,
\label{factorized}
\eea
where as shown on the left, the ``physical" cross section
$\sigma$ depends generically
on a hard scale $Q$ and on a wealth of soft scales,
denoted collectively by $m$.   
The soft scales include
in general the gluon mass, which is zero, as well as
various quark masses, and the scale of the perturbative
coupling, $\Lambda_{\rm QCD}$, encoded in the
running coupling, Eq.\ (\ref{alpharun}).

On the right of Eq.\ (\ref{factorized}), 
we give the schematic factorized form of $\sigma$, in
which the $Q$-dependence and $m$-dependence are
separated.   There is a 
short-distance function $\omega_{\rm SD}$, which is
infrared safe, and a long-distance function $f_{\rm LD}$, 
which for hadronic initial states is not calculable in perturbation theory.
The short- and long-distance functions
are linked by a convolution, denoted
$\otimes$.   For deep-inelastic
or hadron-hadron scattering
the convolution is in partonic momentum fractions, ``$x$",
which is transformed into a simple product by
the moments leading to Eq.\ (\ref{fNevol}) above.   

Dimensional analysis requires that we introduce a new scale,    
$\mu$, the {\it factorization scale}, so that 
$\omega_{\rm SD}$ and $f_{\rm LD}$ can be nontrivial functions of
$Q$ and $m$, respectively.   In effect, the factorization scale
marks the boundary between
short-distance and long-distance dependence.

As indicated in Eq.\ (\ref{factorized}),
factorization is not normally an exact result, but it often
holds up to corrections that behave as inverse
powers of $Q$.   For many important examples, such
as unpolarized deep-inelastic scattering cross sections, corrections
enter only as  $1/Q^2$, and are negligible for many purposes
once $Q$ reaches several GeV. 

In the most familiar examples,
including the structure functions $F_1$ and $F_2$ in (\ref{Tmunudef}),
 the 
$f_{\rm LD}$ are parton distributions, and
we shall refer to them as such.   
The parton distributions themselves can
be expressed in terms of expectation values \cite{CS82} in
hadronic states that fix light-cone components
of the momenta of the partons in question.
We take the light-cone components for any vector $v^\mu$
as  $v^\pm=(1/\sqrt{2})(v^0\pm v^3)$,
with $v^2=2v^+v^--v_T^2$.   A vector whose
only nonvanishing component is $v^+$ or $v^-$
is light-like, $v^2=0$.

For example,  the (spin-averaged) distribution of quark $q$ in nucleon $N$
with momentum $p^\mu=p^+\delta_{\mu +}$, and 
spin $s$, is
\bea
  f_{q/N} (x, \mu^2) &=& {1\over 2}
\sum_s \int^\infty_{-\infty}  {d y^-
\over 2\pi} e^{-i x p^+ y^- }
\,
<  N(p, s) \mid \bar{q} (0^+, y^-, {\bf 0}_\perp)
\nonumber\\
&\ & \hspace{15mm} \times\;
\, {1\over 2}  \gamma^+\; \Phi_n(y^-,0)\;
  q (0) \mid  N(p, s) >\, .
\label{phiqazero}
\eea
We can compare this form to the matrix element
for currents, Eq.\ (\ref{Tmunudef}).
In this case, the factorization scale, $\mu$,
enters because we must renormalize the product
of quark fields that are separated by a light-like distance $y^-$
in the minus direction.   
The operator $\Phi(y^-,0)$ is a ``gauge link",
between the two fields, whose purpose is to render the 
matrix element gauge invariant, and which is defined by
\bea
\Phi_n(y^-,0) =
P \exp \left[-  i g \int_0^{y^-} d l\;
n\cdot A (l n^\mu) \right]\, ,
\eea
with $n^\mu=\delta_{\mu -}$.
Here the field $A^\mu = \sum_a A^\mu_a T_a$ is given as
a matrix in terms of the relevant generators of SU(3),
which for quarks are in fundamental ($3\times 3$) representation.
The symbol $P$ denotes an ordering of these
color matrices along the path between $l=0$ and $l=y^-$.

\subsection{From factorization to evolution}

If we can factorize a cross section as in Eq.\ (\ref{factorized}),
its $Q$-dependence is calculable.  As such, we can
compute it systematically in extensions of the standard model
that include new heavy states, which modify the short-distance
behavior of the theory.   ``New physics", then, is embedded
in a calculable fashion in  $\omega_{\rm SD}$.
While not calculable in perturbation theory, the functions $f_{\rm LD}$
are ``universal", portable from one factorizable process to another.

The key to the portability of parton distributions,
is their ``evolution", which enables us to compute
their dependence on the factorization scale \cite{DGLAP}.
Calculable evolution is not a separate assumption,
but rather a direct consequence of the factorization
in Eq.\ (\ref{factorized}).
We need only observe that the physical cross section
cannot depend on the factorization scale,
\bea
0=\mu{d\over d\mu} \ln \sigma_{\rm phys}(Q,m)\, .
\eea
We can thus separate dependence on $Q$ and $m$ by
requiring that the $\mu$-dependence of the short-distance
function cancel that of the long-distance function,
\bea
\mu{d \ln \omega_{\rm SD} \over d\mu}= - P(\as(\mu)) = - \mu{d \ln f_{\rm LD}\over d\mu}\, .
\label{evolution}
\eea
The ``separation constant" $P(\as)$ can depend only on those
variables that the short- and long-distance functions hold
in common: the coupling and the convolution variables.
Eq.\ (\ref{evolution}) is an evolution
equation.   We can solve it to relate parton distributions
at one $\mu$ to another, and therefore, because we can
always choose $\mu=Q$ in Eq.\ (\ref{factorized}), we can
relate the cross section at one $Q$ to that at another scale, up to
corrections associated with the expansion of $\omega_{\rm SD}(Q/\mu,\as(\mu))$
in the strong coupling.   Of course, this analysis requires that 
$\as$ remain small in the range over which we wish to 
evolve.   Schematically, then, we can exhibit the 
cross section's dependence on
the momentum transfer as
\bea
\sigma_{\rm phys}(Q,m) =  \omega_{\rm SD}(1,\alpha_s(Q))\, 
\otimes\, \exp\left\{  -\; \int_{Q_0}^Q {d\mu'\over \mu'} 
P\left( \alpha_s(\mu')\right) \right\} f_{\rm LD} (Q_0,m)\, ,
\nonumber\\
\eea
just as in Eq.\ (\ref{fNevol}) for the moments of
structure functions.

\subsection{The pattern of a factorized cross section}

A large class of hadronic cross sections can be factorized,
as long as they are defined in a manner consistent with the
energy flow interpretation described in the previous section.
This involves, in general, observing a jet-like structure in
the final state and summing over soft radiation between
the jets. 
The general structure of any such
observable falls into a  form that can be
represented schematically as
\bea
{d\sigma_{a+b\rightarrow {\rm jets\, \{i\}}}(Q) \over dQ}
=  f_{a'/a} \otimes f_{b'/b}\, \otimes {H^{a'+b'\rightarrow \{d_i\}}_{IK}} 
   \times {S^{a'+b'\rightarrow \{d_i\}}_{KI}} \times  {\prod_{{\rm jets}\, \{i\}} J_{d_i}}\, .
   \nonumber\\
  \label{hadfact}
\eea
We can think of this expression as recounting
a (quantum-mechanical) story:  evolved incoming partons 
represented by $f_{a'/a}$, $f_{b'/b}$
 collide and exchange momenta at short distances.   A
 function $H_{IK}$ describes quantum corrections at that scale ($Q$), where 
 $K$ and $I$ identify color exchange in the amplitudes and their 
 complex conjugates, respectively.   These indices are in a color tensor
 basis that  reflects
 the numbers and color representations of all the ``active" partons,
 $a'$, $b'$, and $\{d_i\}$ \cite{colorflow}.
 In general, the color exchange
 at short distances influences the development of the system at
 long distances, through a color-exchange-dependent soft
  function, $S_{KI}$ describing the production of soft particles.   
 Finally, the production of energetic particles and jets is described by
 a set of functions $J_{d_i}$, each 
 specifying the fragmentation of the parent parton $d_i$ of jet $i$.   
 These fragmentation processess
  are mutually incoherent, with a universal evolution into the final states 
  that is itself the result of this factorization.

 Eq.\ (\ref{hadfact}) holds in general to all powers of the coupling, with
 power corrections in hard scales.   The latter, however, can be quite
 complex, involving ratios of the maximum soft energy to
 jet energies: $E_{\rm soft}/E_{\rm jet}$, but also inverse powers
 of the energy of soft radiation.   That is, we also anticipate 
 ``power corrrections" of the form $m/E_{\rm soft}$, with $m$ any
 of the long-distance mass scales in the theory \cite{JETPC}.
 On the perturbative level, the very presence of a
 factorization involving soft, jet and short-distance functions
 ensures more elaborate evolutions, involving
 double-logarithmic corrections \cite{Sudresum}.
 
 It is worth noting that the original cosmic ray jets were not
 of this sort.   Their properties are not computable
 in quite the same way, because for the most part they lack
 truly high-momentum transfer subprocesses, represented
 by $H_{KI}$ in Eq.\ (\ref{hadfact}).   For recent applications
 of perturbative QCD to such ``inclusive" proton-nucleus and
 nucleus-nucleus cross sections, see Ref.\ \cite{Abreu:2007kv}.
 
\begin{figure}[h]
\centerline{\epsfxsize=6cm \epsffile{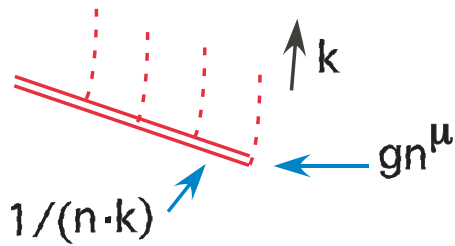} }
\caption{Graphical representation of the ordered exponential, Eq.\ (\ref{oe}).\label{oefig}}
\end{figure}

 A generalization of Eq.\ (\ref{hadfact}) applies in
 hadronic scattering to high-transverse momentum ($p_T$) single-particle
 inclusive cross sections.   
 In this case, the jet functions of Eq.\ (\ref{hadfact}) are
 replaced by fragmentation functions,
 \bea
d\sigma_{A+B\to H+X}(p_T) &=& \sum_c d\bar\sigma_{A+B\to c+X}(p_T/z\mu) \otimes
D_{H/c}(z,m_c,\mu) \nonumber\\
&\ & \hspace{20mm} + {\mathcal O}(m_c^2/p_T^2)\, ,
\eea
with a sum over fragmenting partons $c$.
The cross section $d\bar\sigma$ includes parton distributions 
for the initial state.
 Here, following the formalism developed by Collins and Soper \cite{CS82}, 
 the fragmentation function, $D_{H/c}$ can be defined as 
 a vacuum expectation value similar to those above
 for the distributions, but now involving creation and annihilation
 operators for the observed hadron.   For a gluon to fragment to hadron $H$,
 for example, the function is \cite{CS82}
\bea
D_{H/g}(z,m_c,\mu) &=& -~\frac{1}{16(2\pi)P^+}{\mathrm Tr}_{\mathrm color}
\int dy^-{\mathrm e}^{-i(P^+/z)y^-}\nonumber\\
&\ & \hspace{-25mm}
\times 
\langle 0|G^{+\lambda}(0)\, {[\Phi_-^{(adj)}(0)]^\dagger}\, a_H^\dagger(P^+)\, a_H(P^+)\,
{\Phi_-^{(adj)}(y^-)}\, G^+{}_\lambda(y^-)|0\rangle\, ,
\eea
with $G^\mu{}_\nu$ the gluon field strength and $a_H^\dagger$ the creation
operator for hadron $H$.
 The relevant ordered exponential, or gauge link, for this process is
\bea
\Phi_-^{(adj)}(x^-) = P \exp\left[ -ig\int_0^\infty n\cdot A^{(adj)}\left( (x^-+\lambda)n\right)\, \right]\, ,
\label{oe}
\eea
where $n^\mu$ is a lightlike vector in the opposite direction to the jet.
For gluon fragmentation, the gauge field $A^{(adj)}_\mu$ is an $8\times 8$ matrix in the adjoint
representations of SU(3) generators.
Such a gauge link gives rise to a nice set of diagrammatic
rules, in terms of ``eikonal lines"  in  the $x^-$ direction $(n^\mu=\delta_{\mu -})$,
with vertices $-ign^\mu\times(group\ factors)$ and (linear) propagators $i/(n\cdot k)$,
illustrated in Fig. \ref{oefig}.
To the jet, as it fragments, all that's left of the rest of the world is a gluon
 source moving in the opposite direction, 
 whose entire influence is summarized by the eikonal line.
Similar considerations apply to the parton distributions, Eq.\ (\ref{phiqazero}).

\subsection{The classical basis of factorization}

Where do factorized cross sections like Eq.\ (\ref{hadfact}) come
from?   In the following we review an argument based on
the classical Lorentz transformation properties of fields,
and point out the subtleties of gauge fields in particular \cite{BasuRS}.
An argument based on a classical picture isn't  far-fetched,
precisely because, as noted above, the correspondence principle is the key
to the origin of infrared divergences.
Any accelerated charge must produce classical radiation,
and infinite numbers of soft gluons are required 
to make a classical field.   Thus the classical field has
a lot to tell us about the radiation of soft partons.

\begin{figure}
\centerline{\hskip 3.5 in {\epsfig{figure=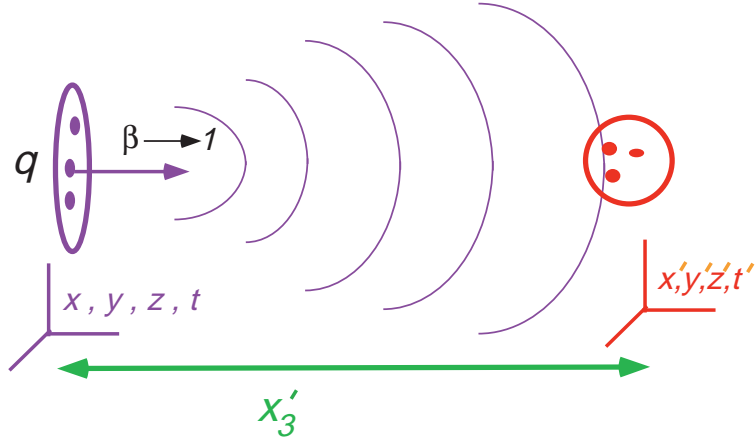,width=1.40 \textwidth }}}
\caption{Schematic representation of the field of an oncoming particle. \label{oncoming}}
\end{figure}

Having said this much, we consider the situation
illustrated in Fig.\ \ref{oncoming}, in which one ``bound state",
approaches another (from the left in the figure) at a relativistic
velocity $\beta\rightarrow 1$ in the $x_3$-direction, carrying with it various point charges,
its ``partons".   The coordinates of the 
bound state on the left are indicated by unprimed 
variables, those on the right by primed.   We will
refer to the former as the projectile, the latter as the target.

Suppose the partons of our projectile are sources for a massless scalar field, 
whose magnitude we denote by $\phi$.  In their own rest frames,
the sources produce a simple $1/|\vec{x}|$ potential.
By definition, the magnitude of a scalar field at any point in space-time
is independent of the coordinate system in which it is observed.
We can thus start with an expression for our massless scalar field
in the rest frame of the projectile, and simply reexpress it
in terms of the coordinates of the target.   To do  so we use
$x_3 = \gamma(\beta ct'-x'_3) \equiv \gamma \Delta'$,
where as usual $\gamma=(1-\beta^2)^{-1/2}$.   
This gives
\bea
 {\phi(x)\ =\ {q \over \sqrt{x_T^2+x_3^2}}} 
\ =\ \phi'(x')\ =\ {q
\over (x_T^2+{\gamma^2}\Delta'{}^2)^{1/2}}\, ,
\eea
where $q$ is a charge and $x_T$ the distance of closest approach, which
is transverse to the motion.
Naturally, the field is maximized in the target coordinates at the time of 
 closest approach, where $\Delta'=0$, that is, at $t' = \frac{1}{\beta c}x'_3$.
At this target time, the magnitude of the field is simply $q/x'_T$.
At all other values of the time $t'$, however, the field of the oncoming
projectile partons is proportional to an explicit factor of 
$1/\gamma$.   In summary,
 the scalar field transforms ``like a ruler", that is,
{at any fixed 
$\Delta' \ne 0$,
the field decreases like $1/\gamma=\sqrt{1-\beta^2}$} as $\gamma\rightarrow \infty$.
This is to say that for any fixed time in the target frame
before closest approach, the field of the
projectile decreases rapidly as the velocity of
the projectile approaches the speed of light.   
This is just a consequence of  length contraction
in elementary special relativity.
When an observer riding on the projectile (!)
measures a distance $x_3$,  then an observer sitting on the target measures a
much larger distance.

Next, we suppose that the sources of the projectile couple to 
the electromagnetic field instead of a scalar field, 
producing in their own rest frames the same $1/|\vec{x}|$
potential, but now as the zeroth component of the vector $A^\mu(x)$.
The following array compares gauge fields to scalar fields from this point
of view.   We compare, on the one hand, the $A^0$ component of the field
in the projectile frame to the $A^0$ component in the target frame,
found by Lorentz transformation, and on the other hand the 
longitudinal (third) component of the {\em electric field} in both frames,
\begin{eqnarray}
 \begin{array}{ccc}
      \underline{{\rm field}} &  \underline{x\ {\rm frame}} &  
\underline{x'\ {\rm frame}} \vspace{8mm} \\
                 {\rm scalar}  &  {q \over |\vec x|} & {q
\over (x_T^2+{\gamma^2}\Delta^2)^{1/2}}
\vspace{8mm} \\
     {\rm gauge} & A^0(x) = {q \over |\vec x|} &  A'{}^0(x')
={q{\gamma}\beta
\over (x_T^2+{\gamma^2}\Delta^2)^{1/2}}
\vspace{8mm} \\
{\rm field\ strength}  & \hspace{10mm}E_3(x)={- q \over |\vec
x|^2} 
\hspace{15mm} & E_3'(x') =
{-q{\gamma}{\Delta} \over (x_T^2+{\gamma^2}\Delta^2)^{3/2}}\, .\\
\ \\
                    \end{array}
\nonumber
\end{eqnarray}
We can ask the same question of the electromagnetic potential
and field strength that we posed for the scalar field: at a fixed target time
before the point of closest approach, how does the field
observed at the target depend on the velocity of the projectile?
The answers for the gauge field and the field strength are
strikingly different.   The gauge potential is actually independent
of $\gamma$ as $\beta\rightarrow 1$ for any fixed $\Delta'\ne 0$!
The vector potential (at least its time component) is not
contracted at all.   On the other hand, the field strength,
as represented by $E_3$, decreases as $1/\gamma^2$, 
which is a much more rapid decrease than even
the scalar field.   

These two behaviors are, of course, consistent, and are reconciled
by the realization that the vector field of a relativistic charge approaches
a total derivative in the target (primed) frame as $\beta\ra 1$,
\bea
A'\, {}^\mu(x') = q \frac{\partial}{\partial x'_\mu}\ \ln \left( \beta c t' \, - \, x'_3\right)
+
{\cal O}( 1 - \beta)\, .
\label{oneminusbeta}
\eea
 The bulk of this  $A^\mu$
  is actually an unphysical polarization, and can be removed by a gauge transformation.
 In contrast, the physical ``force" field $\vec{\bf E}$ of the projectile does  not overlap
the target until the moment of closest approach.
 ``Advanced" effects in the electric field are corrections to the total derivative in Eq. (\ref{oneminusbeta}),
 and hence are of the size
\bea
1- \beta\ \sim\ \frac{1}{2}\; \left [\sqrt{1-\beta^2}\right ]^2\ \sim\ \frac{m^2}{2\omega'{}^2}\, ,
\eea
where $m$ is the mass of the projectile, and $\omega'$ its energy in the target frame.
 This is a power-suppressed behavior, and a typical
 initial-state correction to factorization.

Factorization expresses this contraction effect.   As the oncoming projectile
approaches the speed of light, the appearance of its field is essentially
instantaneous at the time of closest approach.   The projectile then cannot
affect the internal structure of the target, or vice-versa, and the target's
 internal structure is thus
effectively universal among all projectiles, so long as the latter are sufficiently relativistic.
The initial state structure of the target and projectile can then both be
summarized by multiplicative factors, and these are the parton distributions
of Eq.\ (\ref{hadfact}).

This argument, of course, applies only to initial-state interactions,
signals exchanged before the hard interaction.
For final-state processes to respect factorizations like Eq.\ (\ref{hadfact}),
it is also necessary that we define the observable in a manner
consistent with infrared safety.   In addition, for factorization to hold,
we must require that there be a hard scattering.  Otherwise there is
no well-defined time at which the scattering occurs,
and indeed no sharp distinction between the initial state and the final state.
If there is a well-defined hard
scattering, however, low-momentum transfers after that scattering are too late 
to affect the  large momentum
transfer process(es), such as the creation of jets or
of heavy particles.
Similarly, the fragmentation of partons into jets
of hadrons is too late to know details of
the hard scattering, leading to the factorization of fragmentation functions.

\subsection{Factorization in perturbation theory}

Perturbative arguments for factorization in QCD \cite{FactPrf,CSS89} are, unfortunately,
much more complex than the simple classical pictures above.
Nevertheless, the physical observations we have just made
have a direct correlation in perturbation theory, which is worth
pointing out.   We consider a soft gluon,
of momentum $k$ emitted by a fast quark, whose momentum 
 $p^\mu$ is on-shell $(p^2=m^2)$ just after
 this interaction.   In perturbation theory, this will be associated with a factor like
\bea
\bar{u}(p)\; (-ig\, \gamma^\mu\, )\; \frac{ \rlap{\it p}{/} + \rlap{\it k}{/}\; +\; m }{ (p+k)^2\ -\ m^2} 
\ =\ \bar{u}(p)\; (-ig\, )\; \frac{p^\mu}{p\cdot k}\ +\  (IR\ finite)\, ,
\nonumber\\
\eea
where in the second form we have used the Dirac equation,
and have suppressed terms proportional to $k$, which
are infrared finite, as well as color factors.
 In an arbitrary perturbative diagram, the vector $p^\mu$ 
 on the right-hand side will be contracted with the
 propagator of the soft gluon that carries momentum $k$.     
Now suppose we were to choose a gauge for which $p\cdot A=0$, 
in which case the gluon propagator is given (with $p^2=0$) by
\bea
G^{\nu\mu}(k)
=
-\, \frac{i}{k^2}\; \left( g^{\nu\mu}\ - \frac{p^\nu\, k^\mu \, +\, k^\nu\, p^\mu}{p\cdot k} \, \right)\, .
\eea
In this gauge, the soft gluons decouple from the quark.
This argument can easily be generalized beyond lowest
order, and applies to the entire set of collinear partons, whether quark, antiquark
or gluon, in that jet.    No gauge choice like this, of course,
can decouple soft gluons from more than one jet at a time.   But the 
existence of such a gauge for each jet implies that
soft gluon couplings cannot resolve more than the direction
and overall color of a jet \cite{CSS89,Tucci:1984yp}.   This
 is the origin of the ``universality" of soft gluon interactions, 
 and their summary in terms of eikonal lines like those
 of Eq.\ (\ref{oe}) and Fig.\ \ref{oefig},
 which  play a central role in factorization for perturbative QCD.   

\section{Conclusion}

We have summarized a few of the major results of perturbative
QCD, which underly the basic applications of the theory to
hadron-hadron and hadron-lepton collisions at large momentum
transfer.    We have
presented justifications wherever they can be found, in
both classical and quantum intuition.

The coming decade will see unprecedented applications of
the  ideas and methods of
perturbative QCD at the Large Hadron Collider, in proton-proton,
 proton-nucleus and nucleus-nucleus experiments.    Whether as a pesky background
to new physics searches, or as a subject of interest in its own right,
QCD, with its self-generated scales and evolving degrees of
freedom, will remain a benchmark for our understanding
of physics at its most challenging.

\end{document}